\documentclass[twoside,twocolumn,9pt]{article}
\usepackage{extsizes}
\usepackage[super,sort&compress,comma]{natbib} 
\usepackage[version=3]{mhchem}
\usepackage[left=1.5cm, right=1.5cm, top=1.785cm, bottom=2.0cm]{geometry}
\usepackage{balance}
\usepackage{mathptmx}
\usepackage{sectsty}
\usepackage{graphicx} 
\usepackage{lastpage}
\usepackage[format=plain,justification=justified,singlelinecheck=false,font={stretch=1.125,small,sf},labelfont=bf,labelsep=space]{caption}
\usepackage{float}
\usepackage{fancyhdr}
\usepackage{fnpos}
\usepackage[english]{babel}
\addto{\captionsenglish}{%
  \renewcommand{\refname}{Notes and references}
}
\usepackage{array}
\usepackage{droidsans}
\usepackage{charter}
\usepackage[T1]{fontenc}
\usepackage[usenames,dvipsnames]{xcolor}
\usepackage{setspace}
\usepackage[compact]{titlesec}
\usepackage{hyperref}

\usepackage{epstopdf}

\definecolor{cream}{RGB}{222,217,201}

\begin{document}

\pagestyle{fancy}
\thispagestyle{plain}
\fancypagestyle{plain}{
\renewcommand{\headrulewidth}{0pt}
}

\makeFNbottom
\makeatletter
\renewcommand\LARGE{\@setfontsize\LARGE{15pt}{17}}
\renewcommand\Large{\@setfontsize\Large{12pt}{14}}
\renewcommand\large{\@setfontsize\large{10pt}{12}}
\renewcommand\footnotesize{\@setfontsize\footnotesize{7pt}{10}}
\makeatother

\renewcommand{\thefootnote}{\fnsymbol{footnote}}
\renewcommand\footnoterule{\vspace*{1pt}%
\color{cream}\hrule width 3.5in height 0.4pt \color{black}\vspace*{5pt}} 
\setcounter{secnumdepth}{5}

\makeatletter 
\renewcommand\@biblabel[1]{#1}            
\renewcommand\@makefntext[1]%
{\noindent\makebox[0pt][r]{\@thefnmark\,}#1}
\makeatother 
\renewcommand{\figurename}{\small{Fig.}~}
\sectionfont{\sffamily\Large}
\subsectionfont{\normalsize}
\subsubsectionfont{\bf}
\setstretch{1.125} 
\setlength{\skip\footins}{0.8cm}
\setlength{\footnotesep}{0.25cm}
\setlength{\jot}{10pt}
\titlespacing*{\section}{0pt}{4pt}{4pt}
\titlespacing*{\subsection}{0pt}{15pt}{1pt}

\fancyfoot{}
\fancyfoot[LO,RE]{\vspace{-7.1pt}\includegraphics[height=9pt]{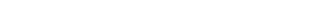}}
\fancyfoot[CO]{\vspace{-7.1pt}\hspace{13.2cm}\includegraphics{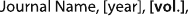}}
\fancyfoot[CE]{\vspace{-7.2pt}\hspace{-14.2cm}\includegraphics{head_foot/RF}}
\fancyfoot[RO]{\footnotesize{\sffamily{1--\pageref{LastPage} ~\textbar  \hspace{2pt}\thepage}}}
\fancyfoot[LE]{\footnotesize{\sffamily{\thepage~\textbar\hspace{3.45cm} 1--\pageref{LastPage}}}}
\fancyhead{}
\renewcommand{\headrulewidth}{0pt} 
\renewcommand{\footrulewidth}{0pt}
\setlength{\arrayrulewidth}{1pt}
\setlength{\columnsep}{6.5mm}
\setlength\bibsep{1pt}

\makeatletter 
\newlength{\figrulesep} 
\setlength{\figrulesep}{0.5\textfloatsep} 

\newcommand{\topfigrule}{\vspace*{-1pt}%
\noindent{\color{cream}\rule[-\figrulesep]{\columnwidth}{1.5pt}} }

\newcommand{\botfigrule}{\vspace*{-2pt}%
\noindent{\color{cream}\rule[\figrulesep]{\columnwidth}{1.5pt}} }

\newcommand{\dblfigrule}{\vspace*{-1pt}%
\noindent{\color{cream}\rule[-\figrulesep]{\textwidth}{1.5pt}} }

\makeatother

\twocolumn[
  \begin{@twocolumnfalse}
{\includegraphics[height=30pt]{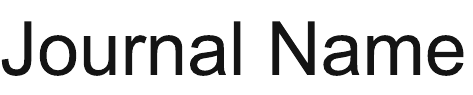}\hfill\raisebox{0pt}[0pt][0pt]{\includegraphics[height=55pt]{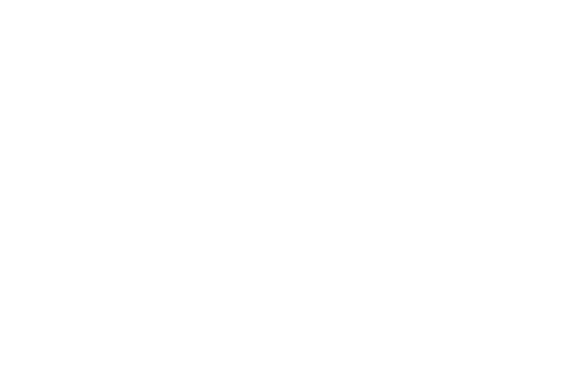}}\\[1ex]
\includegraphics[width=18.5cm]{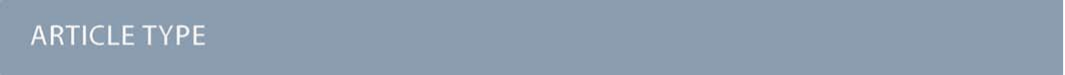}}\par
\vspace{1em}
\sffamily
\begin{tabular}{m{4.5cm} p{13.5cm} }

\includegraphics{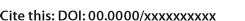} & \noindent\LARGE{\textbf{Biexcitons in Ruddlesden–Popper Metal Halides Probed by Nonlinear Coherent Spectroscopy}} \\
\vspace{0.3cm} & \vspace{0.3cm} \\

 & \noindent\large{Katherine~A.~Koch, \textit{$^{a}$} Carlos~Silva-Acu\~{n}a\textit{$^{b}$}}, and Ajay~Ram~Srimath~Kandada,\textit{$^{a}$}  \\

\includegraphics{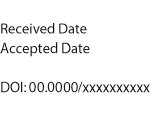} & \noindent\normalsize{Excitons and their correlated complexes underpin the rich photophysics of quantum-confined semiconductors. Among these, biexcitons -- bound states of two electrons and two holes -- provide a sensitive probe of Coulomb correlations, exciton-exciton interactions, and the role of the dielectric environment. In Ruddlesden–Popper metal halide materials (RPMHs), strong quantum and dielectric confinement stabilize excitons with binding energies of hundreds of meV, creating an ideal platform for multi-exciton phenomena. Conventional linear spectroscopies, such as photoluminescence and transient absorption, reveal biexciton signatures but suffer from spectral congestion and reabsorption artifacts. Two-dimensional coherent spectroscopies, particularly two-quantum (2Q) multidimensional techniques, uniquely access multi-exciton coherences and provide unambiguous estimates of biexciton binding energies. This minireview surveys the spectroscopic evidence for biexcitons in RPMHs, highlights the advantages of nonlinear multidimensional approaches, and situates biexciton physics within the broader context of excitonic materials, including GaAs quantum wells, quantum dots, and transition-metal dichalcogenides. By emphasizing the interplay of exciton-exciton annihilation, excitation-induced dephasing, and biexciton formation, we argue that multidimensional coherent spectroscopy offers the most reliable pathway to disentangle many-body interactions in quantum-well derivatives of metal-halide perovskites.} \\

\end{tabular}

 \end{@twocolumnfalse} \vspace{0.6cm}

  ]

\renewcommand*\rmdefault{bch}\normalfont\upshape
\rmfamily
\section*{}
\vspace{-1cm}


\footnotetext{\textit{$^{a}$~Department of Physics and Center for Functional Materials, Wake Forest University, 2090 Eure Drive, Winston-Salem, NC~27109, United~States; E-mail: srimatar@wfu.edu}}
\footnotetext{\textit{$^{b}$~Institut Courtois, Universit\'e de Montr\'eal, 1375 Avenue Th\'er\`ese-Lavoie-Roux, Montr\'eal H2V~0B3, Qu\'ebec, Canada, E-mail: carlos.silva@umontreal.ca }}




\section*{Introduction}

Excitons --- Coulomb-bound electron–hole pairs --- are central to the optical response of semiconductors. By enhancing absorption and emission cross-sections~\cite{huang2023origin, luo2021regulating, zhao2023pressure}, facilitating energy transport~\cite{wehrenfennig2013high, yi2016intrinsic, raimondo2013exciton, gong2024boosting}, and modulating optoelectronic performance~\cite{liang2022boosting, chen2021approaching, mak2016photonics, mueller2018exciton, xiao2017excitons}, they underpin a wide range of light–matter interactions. In quantum-confined systems, including inorganic nanostructures and hybrid organic–inorganic materials, excitons often exhibit sharp, well-resolved resonances in optical spectra. These atomic-like transitions provide an excellent materials platform for exploring many-body physics through fluence- and temperature-dependent studies of linear and nonlinear optical responses.

Among the manifold excitonic complexes, biexcitons, bound states of two excitons, \textit{i.e}.\ two-electron and two-hole correlations, offer a compelling window into correlated excitation dynamics, exciton–exciton interactions, and the role of the surrounding bath. Their formation, stability, and spectral signatures are highly sensitive to dimensionality, dielectric environment, and quantum confinement, making them ideal probes of fundamental condensed-matter physics and material-specific behaviour~\cite{singh1996binding}. Understanding biexciton properties is not only of fundamental interest, but also critical to the advancement of quantum optoelectronic technologies, including entangled photon sources~\cite{winik2017demand, stace2003entangled, gywat2002biexcitons}, quantum gates~\cite{troiani2000exploiting, li2003all, shojaei2011biexcitonic}, and nonlinear light-harvesting platforms~\cite{wei2023charged, polimeno2020observation, chase1979evidence}.

Biexcitonic coupling fundamentally arises from four-particle Coulomb interactions between the constituting electrons and holes. A similar many-body interaction potential also drives other inter-exciton processes, such as exciton-exciton annihilation~\cite{vosco2025exciton, steinhoff2021microscopic} (EEA) and excitation-induced dephasing~\cite{wang1993transient, srimath2020stochastic} (EID), which are more ubiquitously observed and reported in excitonic materials. 
In general, such many-exciton correlations are consequential. 
EEA describes the non-radiative recombination of excitons, where the energy of one exciton is transferred to another.  It is commonly investigated through fluence-dependent time-resolved photoluminescence (TRPL), where faster recombination rates indicate an increased probability of annihilation at higher carrier densities~\cite{delport2019exciton, deng2020long, yuan2015exciton}. EID arises from the incoherent Coulomb elastic scattering between multiple excitons, giving rise to faster dephasing dynamics. EID can be estimated by measuring the dephasing dynamics via the homogenous linewidth, which is most accurately extracted from 2D coherent electronic spectroscopy (2DES)~\cite{schneider2004excitation, srimath2020stochastic, stone2009exciton, katsch2020exciton}. Biexciton formation corresponds to the creation of a new quasiparticle, comprised of two excitons bound via an attractive or repulsive Coulomb interaction. Generally, the presence of biexciton states is observed through photoluminescence (PL) spectroscopy~\cite{birkedal1996binding, cho2024size, hayakawa2014binding}, where a secondary lower-energy resonance appears with increasing fluence, or through transient absorption spectroscopy by studying the excited-state absorption features~\cite{styers2008exciton, zhang2016understanding, shukla2020effect, sewall2008state, yumoto2018hot}. 

Although each of these processes has been studied extensively, their interplay remains poorly understood. In particular, EEA and biexciton formation represent competing outcomes of exciton-exciton interactions: annihilation funnels excitons into non-radiative loss channels, whereas biexciton formation stabilizes them into a bound correlated state. Disentangling the relative contributions of these processes and understanding the conditions under which one dominates over the other is therefore critical for developing a comprehensive picture of exciton photophysics. A central problem here is the quantification of EEA and biexciton. While EEA is more readily accessible as a fluence-dependent recombination rate, biexcitonic couplings are quantified through the estimation of the biexciton binding energy. The biexciton binding energy, defined as the energy difference between the biexciton state and the unbound exciton state, serves as a representation of the underlying many-body interaction potential. 

Several spectroscopic approaches are used to measure exciton binding energies, including straightforward linear spectroscopy techniques and more intricate measurements using non-linear or multi-dimensional spectroscopy techniques. One of the most widely used techniques is photoluminescence (PL) spectroscopy, where the energy difference between the exciton and biexciton emission peaks provides a direct estimate of the biexciton binding energy~\cite{birkedal1996binding, cho2024size, hayakawa2014binding}. Transient absorption spectroscopy is also used to estimate the biexciton binding energy, where the energy of the photo-induced absorption feature corresponds to the excited state absorption from the exciton to the biexciton state~\cite{styers2008exciton, zhang2016understanding, shukla2020effect, sewall2008state, yumoto2018hot}. While effective in simple systems with well-separated resonances, these techniques encounter limitations in two-dimensional materials, where overlapping spectral features introduce ambiguity. For this reason, increasing emphasis has been placed on nonlinear optical approaches, in particular four-wave mixing techniques. Among these, two-quantum (2Q) multidimensional spectroscopy stands out, as it directly accesses multi-exciton coherences and provides the most accurate measurement of the biexciton binding energy. In what follows, we argue that 2Q multidimensional spectroscopy offers the most reliable probe of biexciton properties. 

In this review, we narrow our focus to the spectroscopic signatures of biexcitons, specifically in Ruddlesden-Popper metal halide perovskites. Two-dimensional (2D) halide perovskites have recently emerged as a powerful platform at the intersection of the systems discussed above, combining the strong excitonic effects of monolayers of transition metal dichalcogenides with the chemical and structural tunability of quantum dots. 2D perovskites are a structural derivative of their 3D counterparts. The elongation of the A-site organic or inorganic cation creates a layered quantum well-like structure, where the octahedral lattice layer is electronically separated by the elongated cation (see structural schematic in Fig.~\ref{fig:linear})~\cite{burgos2020exciton, srimath2020exciton}. Strong dielectric and quantum confinement lead to exciton binding energies on the order of hundreds of meV~\cite{hong1992dielectric, kato2003extremely}. Evidence for biexciton states in 2D perovskites has been obtained through photoluminescence (PL) and transient absorption (TA) spectroscopy~\cite{fang2020band, he2022multicolor, armstrong2023spatial, williams2018energy, li2020biexcitons}, while emerging nonlinear spectroscopies are beginning to reveal their dynamics, coherence properties, and state-specific character with greater precision~\cite{thouin2018stable, koch2025spectroscopic}. In this review, we focus on biexciton signatures in 2D Ruddlesden–Popper metal halides (RPMHs), with particular emphasis on two-dimensional electronic coherence spectroscopy as a uniquely powerful approach for extracting accurate and detailed information about biexciton states.

In the sections that follow, we begin with a concise overview of exciton physics in 
Ruddlesden Popper metal halides (RPMHs), highlighting their distinctive optical signatures and the role of quantum confinement. Since our central aim is to demonstrate that two-dimensional electronic spectroscopy (2DES) offers the most precise approach for estimating biexciton binding energies and unraveling the subtleties of exciton–exciton coupling, we next provide a pedagogical introduction to the experimental technique and the theoretical framework used to analyze spectral lineshapes. Building on this foundation, we present spectroscopic investigations of biexcitons in RPMHs, emphasizing the rich and intricate nature of many-body interactions in these systems. We conclude by comparing biexciton physics in RPMHs with other excitonic platforms, including GaAs quantum wells, quantum dots, and TMDCs, and offer our perspective on the future of biexciton spectroscopy.

\section*{Excitons in RPMHs}

The multiple quantum-well structure in RPMHs introduces significant quantum and dielectric confinement, leading to an enhanced stability of the photogenerated excitons, with binding energies on the order of hundreds of meV~\cite{kim2025exciton, marongiu2019role, blancon2018scaling, straus2016direct, straus2018electrons, hansen2024measuring, dyksik2025steric}. These excitations are confined to the metal-halide octahedral lattice layer and govern the optical and optoelectronic properties of the material. The hybrid lattice, with its soft and dynamic organic-inorganic framework, has a substantial effect on the electronic excitations~\cite{bakulin2015real, zhu2016screening, zhu2016organic}, making exciton-phonon coupling a central area of investigation for understanding the photoexcitation dynamics in RPMHs. 

\begin{figure}[ht]
\centering
\includegraphics[width=\columnwidth]{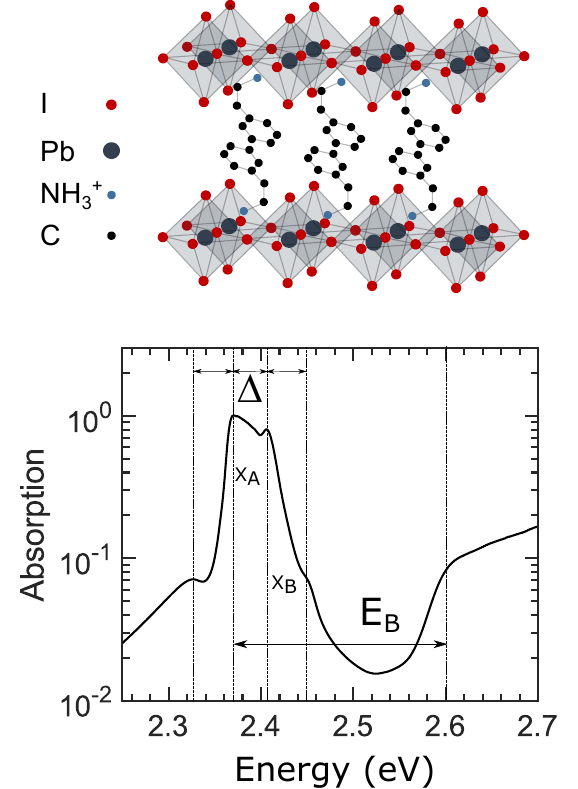}
\caption{Crystal structure of a prototypical 2D HOIP: phenylethylammonium lead iodide (\ce{(PEA)2PbI4}). (Bottom) Linear absorption spectrum of \ce{(PEA)2PbI4} taken at $T= 5$\,K; $\Delta \sim 35\pm 5$\,meV represents the energy spacing within the excitonic fine-structure and $E_B \sim 250$\,meV is the exciton binding energy associated with the main exciton peak. Figure extracted from Ref.~\citenum{srimath2020exciton}. }
\label{fig:linear}
\end{figure}

A distinguishing feature of these materials is the complicated fine structure observed at low temperatures, where multiple resonances appear in the spectra~\cite{dyksik2025steric, posmyk2024bright, ishihara1990optical, goto2006localization, kitazawa2010excitons, kitazawa2010optical, kitazawa2011synthesis, kitazawa2012temperature, shimizu2005influence, shimizu2006photoluminescence}. The nature of the resonances is a highly contested topic in the spectroscopic community, but two main hypotheses prevail: the unique resonances either represent a vibronic progression of a single excitonic transition~\cite{straus2018electrons, straus2016direct}, or they correspond to a family of coexisting excitons with distinct characteristics~\cite{srimath2020exciton, thouin2019phonon, thouin2019enhanced, neutzner2018exciton, biswas2024exciton, fu2021electronic}. Extensive experimental evidence, including work from our own groups~\cite{thouin2019phonon, koch2025structure, srimath2020exciton}, supports the latter picture, where excitations are better described as exciton-polarons --- Coulombically bound electron and hole quasi-particles dressed by lattice vibrations. In this framework, the fine structure reflects a manifold of excitonic states, each distinguished by its degree of lattice dressing. A detailed review of these competing hypotheses is beyond the scope of the present article. Instead, we will focus on multi-exciton spectroscopy in RPMHs by exploring various spectroscopic techniques and comparing the information revealed about the many-body interactions.

Linear spectroscopy, including absorption and photoluminescence, has been the foundational tool for probing excitonic properties in RPMHs. Absorption spectroscopy gives direct access to the intricate fine structure mentioned above, yielding information on excitonic peak energies and their associated oscillator strengths. The band gap energy ($E_g$) can be estimated from the absorption onset (Fig.~\ref{fig:linear}), while fitting procedures based on Elliott analysis may be used to extract exciton binding energies~\cite{neutzner2018exciton, hansen2024measuring, elliott1957intensity, haug2009quantum}. However, this approach is challenging in RPMHs because it relies on the clear visibility of interband continuum features, which are often obscured by strong excitonic absorption. Complementary estimates have been obtained from temperature-dependent PL measurements, where quenching of the free-exciton emission with increasing temperature is fit to an Arrhenius model to determine the ionization energy, taken as the exciton binding energy~\cite{hansen2024measuring, hansen2022low}. Beyond binding energies, PL spectra provide valuable insight into the radiative recombination pathway, defect states (particularly well studied in TMDCs), and many-body dynamics. For example, the appearance of red-shifted spectral shoulders or additional emission features has been attributed to the relaxation of a biexciton state~\cite{birkedal1996binding, cho2024size, hayakawa2014binding, ando2012photoluminescence, you2015observation}, allowing estimates of biexciton binding energies from steady-state measurements.  

\begin{figure*}[h!]
    \centering
    \includegraphics[width=17cm]{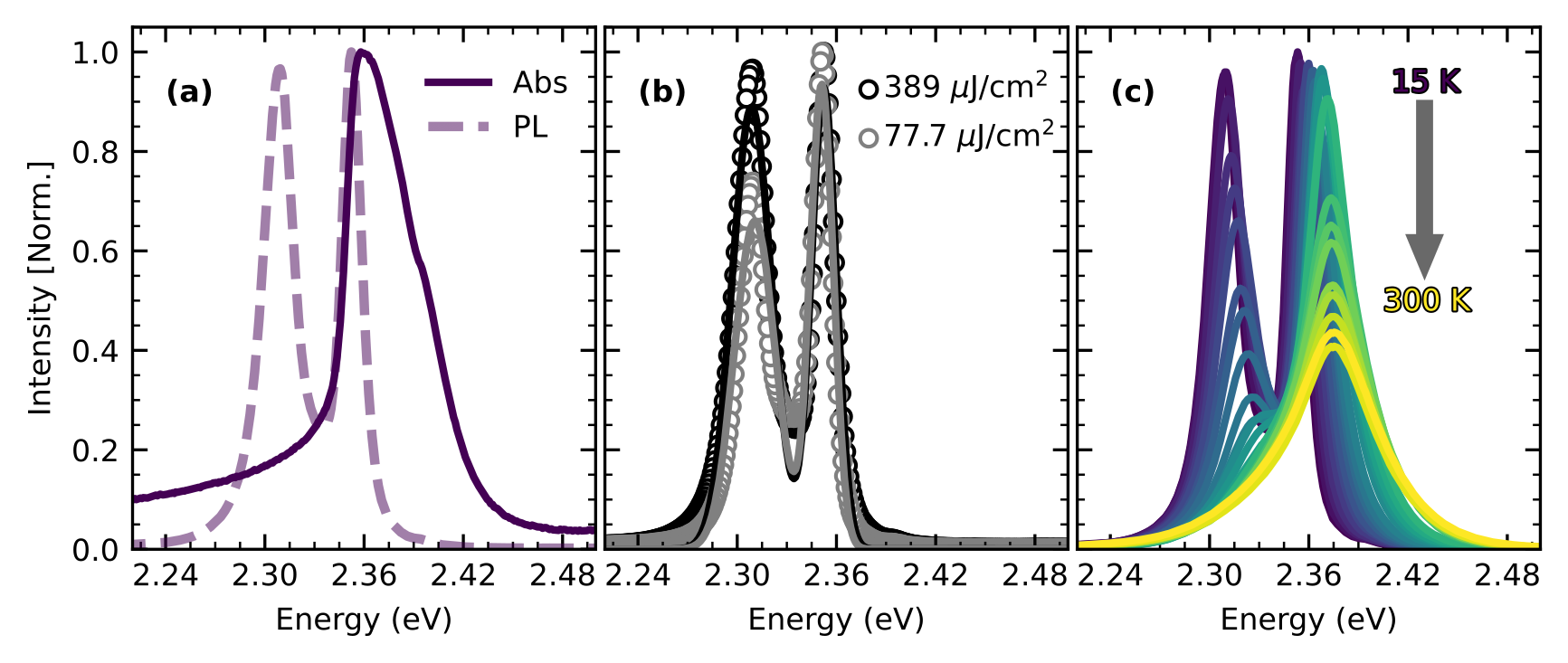}
    \caption{(a) Absorption (solid) and photoluminescence (dashed) spectra, of \ce{(F-PEA)2PbI4} measured at 15 K. (b) Measured photoluminescence spectra (dots) of \ce{(F-PEA)2PbI4} taken at 15 K, fit to a double Gaussian function (lines) to determine peak positions and calculate the biexciton binding energy. (c) Temperature dependent photoluminesence, measured with a pump fluence of 389 $\mu$J/cm$^2$. Figure reproduced from Ref.~\citenum{koch2025spectroscopic}.}
    \label{fig:FPEA_abs_PL}
\end{figure*}


Fig.~\ref{fig:FPEA_abs_PL}(a) shows the linear absorption spectrum measured at 15\,K of the polycrystalline thin film of \ce{(F-PEA)2PbI4}, exhibiting a characteristic excitonic peak accompanied by a broad shoulder at higher energies. The PL spectrum is shown in Fig.~\ref{fig:FPEA_abs_PL}(b), obtained via the photo-excitation of carriers in the continuum. The dominant resonance ($\sim 2.3$5\,eV) in the PL spectra can be assigned to the radiative recombination of excitons, while the red-shifted secondary resonance ($\sim$ 2.30\,eV) has been attributed to relaxation from a biexciton state~\cite{kaiser2021free, goto2006localization, makino2005induced}, based on its fluence dependence. Precise estimates of the peak energies can be obtained by fitting the spectrum to a double Gaussian function, where the difference between the peaks provides the first estimation of the biexciton binding energy, which is 44\,meV, aligning well with previously reported values for RPMHs~\cite{koch2025spectroscopic}. It should be noted, however, that re-absorption effects, particularly in samples with small Stokes shifts such as \ce{(F-PEA)2PbI4}, may lead to an underestimation of the biexciton binding energy.  

A further limitation of linear spectroscopy is its sensitivity to spectral broadening, which becomes increasingly severe at higher temperatures. As shown in Fig.~\ref{fig:FPEA_abs_PL}(c), the temperature-dependent PL spectra exhibits a reduction of the biexciton resonance, and the energy difference between the biexciton and exciton peaks decreases with increasing temperature. While this behaviour may suggest a reduction of the biexciton binding energy with increasing temperature, the broadened spectral features add ambiguity to this observation. Taken all together, these challenges highlight the strengths and limitations of linear methods. They provide the first evidence of biexcitons states and rough binding energy estimates, but cannot definitively disentangle biexciton contributions from overlapping, excitonic, trionic (exciton and a charged particle), or phonon-assisted processes, making them insufficient for a complete analysis of the biexciton photophysics. To overcome these ambiguities, nonlinear and multidimensional spectroscopies are essential as they can isolate multi-exciton coherences and provide the most accurate determination of biexciton properties in RMPHs. 

\section*{Two-dimensional coherent spectroscopy}


Most scientists are familiar with interference patterns, like those in a Michelson interferometer or double-slit experiment. These arise because light waves maintain a fixed phase relationship (i.e., coherence) over time or space. When coherent beams overlap, their electric fields add constructively or destructively depending on their relative phase. This is linear coherence: the interference is between fields, and the system (a screen or a detector) responds linearly to the intensity.

In 2D coherent spectroscopy, we extend this idea ---  instead of light interfering with itself in space, we use phase-controlled ultrafast pulses to create and manipulate quantum coherences within the material. The interference now happens between quantum pathways, which are different sequences of light–matter interactions, and the signal arises from the material’s nonlinear polarization. So while in linear optics the detector sees interference between light fields, in nonlinear coherent spectroscopy the detector sees interference between quantum pathways encoded in the material’s response.

To understand how quantum interference arises in nonlinear spectroscopy, it is helpful to introduce the density matrix, which describes the state of a quantum system in terms of both populations and coherences. The diagonal elements of the density matrix ($|i\rangle \langle i|$) represent populations, which quantify how much of the system resides in a given energy level. The off-diagonal elements ($|i\rangle \langle f|$) encode coherences, or phase relationships between states. These coherences are responsible for quantum interference and are central to the nonlinear optical response. In nonlinear spectroscopy, we control the evolution of the density matrix ($\rho_{ij} = |\psi_i\rangle \langle \psi_j|$) using sequences of light–matter interactions. A single interaction (linear absorption) creates a coherence between ground and excited states. A second interaction can convert that coherence into a population, and subsequent interactions can regenerate or manipulate coherences. The number and timing of these interactions determine whether the system evolves through population dynamics or retains phase coherence—each pathway contributing differently to the emitted signal.

These quantum pathways, each representing a distinct sequence of light–matter interactions, are not only shaped by the timing and phase of the excitation pulses, but also by the phase-matching conditions imposed by the geometry and wavevector configuration of the experiment. In nonlinear optics, phase matching ensures that the generated signal constructively interferes in the sample, allowing coherent buildup of the emitted field. This is a spatial analog to temporal coherence: just as phase-stable pulses allow interference between quantum pathways in time, phase matching ensures that those pathways interfere constructively in space.

By selecting specific phase-matching directions (like rephasing, non-rephasing, or double quantum pathways), we can isolate different components of the nonlinear response, such as ground-state bleaching, excited-state absorption, and two-quantum signatures. Each of these pathways contributes to the overall signal with a distinct phase evolution, and only those that satisfy the phase matching condition will be coherently amplified and detected.

\begin{figure*}[h]
  \centering
  \includegraphics[width=\textwidth]{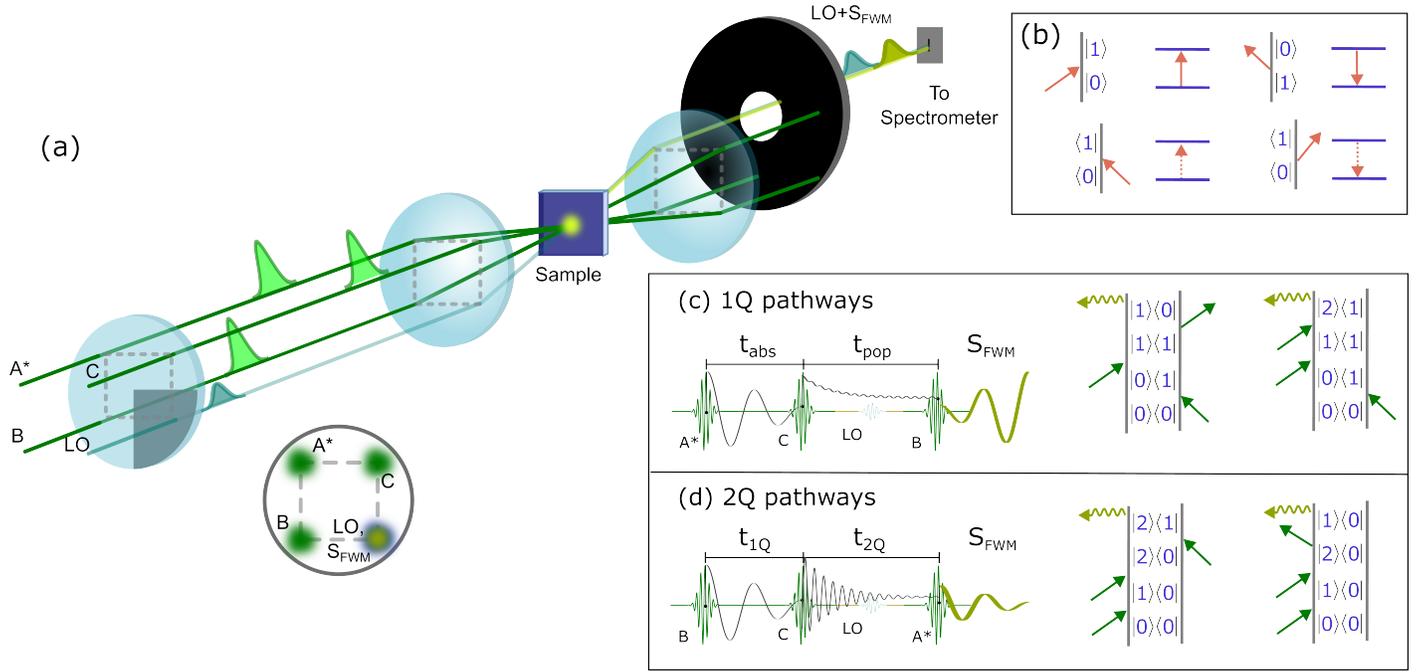}
  \caption{(a) The geometry of the excitation pulse-train beam pattern (blue-green) and the resonant four-wave mixing signal (SFWM, yellow-green), detected by interference with a local oscillator (LO). We use the BoxCARS beam geometry, in which three pulse trains (A$^{\star}$, B, C) propagating along the corners of a square are focused onto the sample with a common lens, defining incident wave vectors $\vec{k}_A$, $\vec{k}_B$, and $\vec{k}_C$. The LO beam, on the fourth apex of the incident beam geometry, co-propagates with $S_{\mathrm{FWM}}$ with the wave vector imposed by the chosen phase-matching conditions. (b) Schematic representation of the effect of light-matter interactions in double-barreled Feynman diagrams. The arrows represent the light-matter interaction, either on the bra (left-hand side) or ket (right-hand side) of the density matrix. Arrows that point to the diagram represent excitation along the ladder of states, while arrows pointing away represent de-excitation. By changing the time-ordering of the pulse sequence in the fixed phase-matching geometry displayed in part (a), we measure two distinct nonlinear responses: (c) 1Q rephasing signal and (d) 2Q non-rephasing signal.}
  \label{fig:pathways}
\end{figure*}


In a typical experiment, three ultrafast pump pulses are used, each characterized by a wavevector $\Vec{k_A}$, $\Vec{k_B}$ and $\Vec{k_C}$. These pulses are arranged in a non-collinear geometry known as the BOXCARS configuration, which ensures spatial separation of the generated signals via phase matching. The pulses are temporally short (approximating delta functions) and interact with the sample in a controlled time-ordered sequence. This sequence induces a third-order nonlinear polarization in the material, which emits a coherent signal in the phase-matched direction: $\Vec{k_{sig}}= -\Vec{k_A}+\Vec{k_B}+\Vec{k_C}$. This signal is detected using heterodyne techniques, which preserve both amplitude and phase information, allowing full reconstruction of the complex-valued response function.

To elucidate the interference between excitation pathways in 2D coherent spectroscopy, we employ a diagrammatic perturbation theory based on the density matrix formalism. This approach offers a compact and intuitive visual representation of how quantum coherences and populations evolve under multiple field interactions. Each interaction is treated as a perturbation, and the sequence of interactions, along with their timing and phase, determines the contribution of a given quantum pathway to the overall signal. The multi-field interaction eventually results in the generation of nonlinear polarization, which emits radiation in the phase-matched direction. \textbf{Emission in the context of 2D spectroscopy refers to the emission of coherent radiation by the time-varying nonlinear polarization $\vec{P}(\vec{r},t)$, induced by the three time-ordered light-matter interactions}. 

Double-sided Feynman diagrams offer a powerful visual language for tracking how quantum states evolve under the influence of ultrafast laser pulses. Unlike more widely used energy-level diagrams, these representations depict the evolution of the system's density matrix. Essentially, it helps us to book-keep the evolution of population and coherence. Each diagram consists of two vertical lines: the left side represents the bra side of the density matrix and the right side the ket of the quantum state. Interactions with light are represented by arrows pointing toward or away from these lines, indicating absorption or emission events. By following the sequence of arrows, one can trace how the system moves through different quantum states -- whether it's coherently oscillating between levels or relaxing into a population.

The direction and placement of each arrow also determine which side of the density matrix the interaction occurs on. This distinction is not merely notational: it directly influences the sign of the wavevector in the phase-matching condition, linking the spatial direction of the emitted signal to the underlying quantum pathway. A negative sign indicates a field acting on the bra side, effectively reversing the coherence evolution relative to a ket-side interaction. 


To illustrate how specific quantum pathways emerge, let us consider a time-ordered sequence where pulse A arrives first, followed by pulse B after a time $t_1$ and then pulse C after $t_2$. Under the phase-matching condition -- $\Vec{k_{sig}}= -\Vec{k_A}+\Vec{k_B}+\Vec{k_C}$, this time ordering selects a class of \textbf{rephasing} diagrams. The first interaction (from pulse A) acts on the bra side of the density matrix, initiating a coherence between the ground and excited state. This coherence evolves during the delay $t_1$, accumulating phase. The second interaction (pulse B) acts on the ket side, converting the coherence into a population. After a waiting time $t_2$, the third interaction (pulse C) reintroduces a coherence that evolves during the detection time and emits the signal. Because the initial and final coherences evolve with opposite phase signs, inhomogeneous broadening is partially canceled, producing an echo-like signal. The nonlinear signal here is measured by scanning $t_1$, which is then Fourier transformed to yield a one-quantum (1Q) correlation map, where diagonal peaks reflect the ground-state bleach signatures of the transitions, and off-diagonal peaks reveal couplings between distinct states. 

Now consider a different time-ordering: pulse B arrives first, followed by pulse C, and then pulse A. With the same phase-matching condition, this sequence isolates double-quantum (2Q) pathways. The first two interactions (B and C) both act on the ket side, driving the system into a coherence between singly and doubly excited states. This 2Q coherence evolves during the waiting time $t_2$, encoding multi-exciton correlations and anharmonicities. The third interaction (pulse A) acts on the bra side, converting the 2Q coherence into a 1Q coherence that emits the signal. The resulting spectrum, which is obtained by Fourier transforming along $t_2$, maps the energy and coupling of doubly excited states, offering insight into many-body interactions and state mixing that are inaccessible via 1Q pathways alone.

\section*{Biexcitons in RPMHs}

\begin{figure*}[h]
    \centering
    \includegraphics[width=\textwidth]{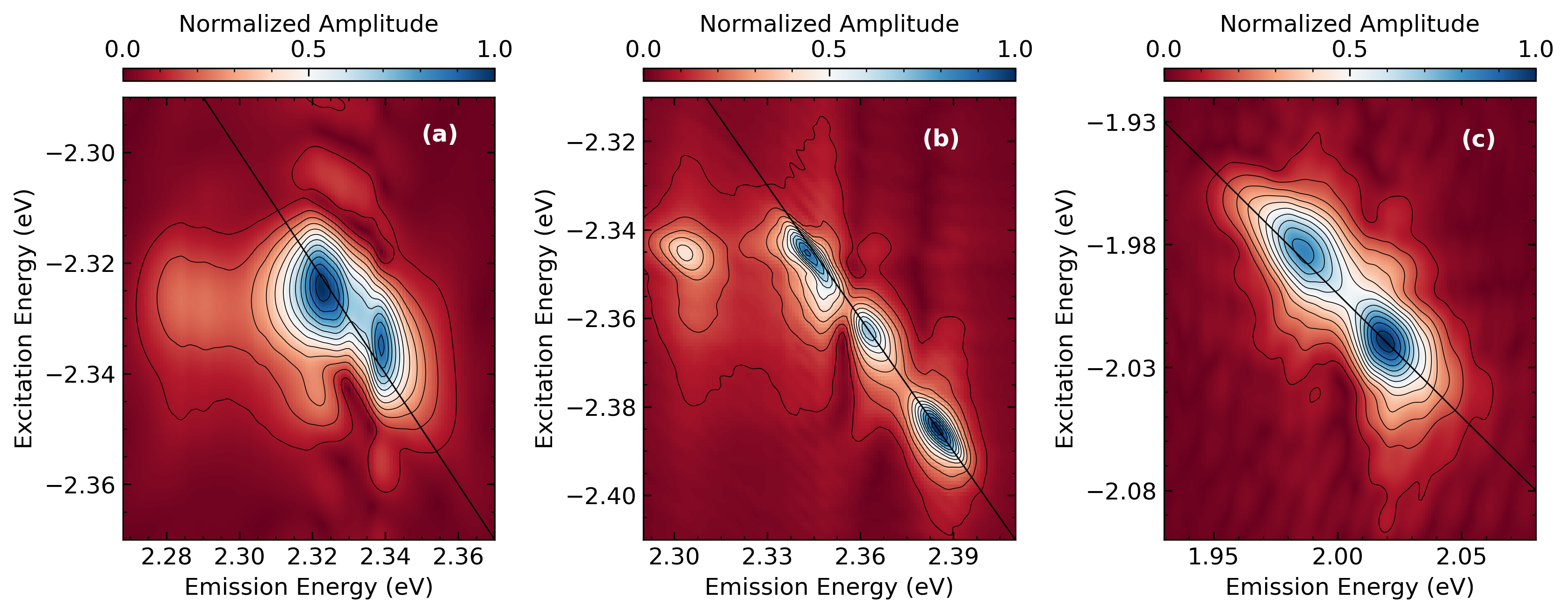}
    \caption{Absolute one-quantum rephasing 2D coherent spectrum of (a) \ce{(F-PEA)2PbI4} measured at 7K (Data reproduced from Ref.~\citenum{koch2025spectroscopic}), (b) \ce{(PEA)2PbI4} measured at 5\,K, (Data reproduced from Ref.~\citenum{thouin2019enhanced}) and (c) \ce{(PEA)2SnI4} measured at 5\,K (Data reproduced from Ref.~\citenum{rojas2023many}).}
    \label{fig:1Q}
\end{figure*}

With a working understanding of 2DES, we now turn to the one-quantum (1Q) re-phasing spectra of RPMHs. Figures~\ref{fig:1Q}(a)-(c) show the absolute and real components of the 2DES-1Q spectrum for \ce{(F-PEA)2PbI4}, \ce{(F-PEA)2PbI4}, and \ce{(F-PEA)2PbI4}, respectively. In this experimental scheme, the first pulse generates a coherence, the second creates a population, and the third pulse induces a time-varying third order polarization that radiates the signal field at $\Vec{k}_{sig}=-\Vec{k}_{a}+\Vec{k}_{b}+\Vec{k}_{c}$, see Fig.~\ref{fig:pathways}(c). For these measurements, the pump laser spectrum is tuned to cover all excitonic features present in the linear absorption spectra of the respective films~\cite{koch2025spectroscopic, thouin2018stable, rojas2023many}. A major advantage of 2DES over linear spectroscopy lies in its ability to disentangle homogenous and inhomogeneous broadening. The 2D coherent lineshape reflects the inhomogeneous Gaussian spectrum along the diagonal, while the homogeneous Lorentzian spectrum is along the antidiagonal~\cite{srimath2022homogeneous}. Focusing on \ce{(F-PEA)2PbI4}, it is evident in Fig.~\ref{fig:1Q}(a) that two distinct resonances, $X_1$ and $X_2$, emerge along the diagonal. These features are obscured in the linear absorption due to inhomogeneous broadening, but are clearly resolved in the 2DES-1Q spectrum, underscoring the power of multi-dimensional spectroscopy. Similar features can be seen in the 1Q response across all three samples, where three resonances are revealed for \ce{(PEA)2PbI4} and two for \ce{(PEA)2SnI4}, all of which are along the diagonal axis, indicating these transitions share the same ground state. While detailed analysis of the diagonal features lineshapes can reveal exciton–phonon coupling strengths and many-body scattering processes~\cite{rojas2023many, thouin2019enhanced}, such discussion is beyond the scope of this review. Instead, focus on the off-diagonal features, which provide insight into biexcitons and their binding energy. 

The most prominent off-diagonal feature of Fig.~\ref{fig:1Q}(a) and (b), red-shifted relative to the diagonal, is attributed to excited state absorption(ESA) from the 1Q states to higher-lying 2Q transitions. This feature allows for the estimation of the biexciton binding energy, which corresponds to the energy difference between the excitation energy (ground state to exciton transition) and the emission energy of the ESA feature (exciton to biexciton transition). A key challenge in this approach lies in accurately determining the peak position of the ESA feature, as its broad spectral profile complicates straightforward analysis. To address this, gradient-based methods have been employed, generating vector maps that enable more precise localization of the ESA peak. A more detailed description of this approach can be found in Ref.~\citenum{koch2025spectroscopic}, where a biexciton binding energy ($E_{B_1}$) of 50\,meV is reported for \ce{(F-PEA)2PbI4} based on the low temperature 1Q spectra. 

In contrast, the ESA off-diagonal feature in \ce{(PEA)2SnI4} is much less pronounced. Given that the biexciton binding energy in tin-based perovskites is significantly smaller than in their lead-based counterparts, it is reasonable that the ESA feature appears closer to the diagonal. This highlights a limitation of 1Q spectra for probing multi-excitonic interactions: spectral congestion and overlap between diagonal and off-diagonal features make it difficult to accurately resolve biexciton coherences, particularly in systems with small binding energies. To overcome these challenges, one must probe the coherences of multi-quantum states directly, an approach made possible through two-quantum (2Q) 2DES, which provides the most direct and unambiguous identification of biexciton signatures.

\begin{figure*}[ht]
	\centering
	\includegraphics[width=\textwidth]{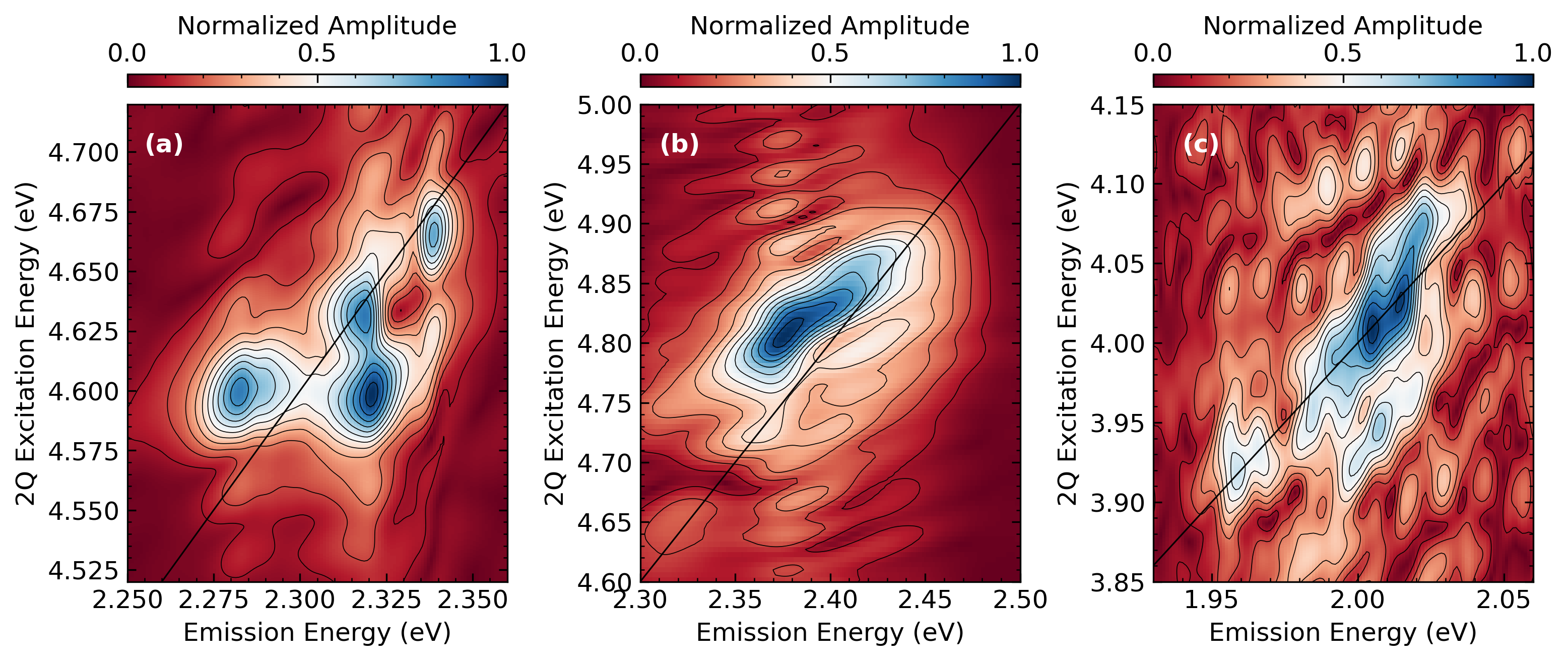}
	\caption{Absolute two-quantum non-rephasing 2D coherent spectrum of (a) \ce{(F-PEA)2PbI4} measured at 7K, (b) \ce{(PEA)2PbI4} measured at 5\,K, and (c) \ce{(PEA)2SnI4} measured at 5\,K. Figure (b) and (c) extracted and modified from Ref.~\citenum{thouin2018stable} and \citenum{rojas2023many}.}
	\label{fig:2Q_fig}
\end{figure*}

A specific pulse sequence, see Fig.~\ref{fig:pathways}(d), enables the direct visualization of double quantum coherences, without passing through a population term. This unique capability makes 2Q-2DES a particularly robust and selective method for probing higher-lying correlated states, such as biexcitons, that are otherwise obscured in conventional linear or one-quantum spectroscopies. The result of such 2Q-2DES measurement is a nonrephasing spectrum that correlates the energy of the 2Q states (2Q Excitation Energy) with the energy of one-quantum excitations (Emission Energy), see Fig.~\ref{fig:2Q_fig}. Typical biexciton signatures of a 2Q-2DES response include a feature along the diagonal, if the two-quantum state has zero binding energy relative to the 1Q transition, but finite binding energy results in features below (above) the diagonal due to attractive (repulsive) Coulomb interactions. The energy offset from the diagonal quantifies the biexciton binding energy of the system. 

In contrast to the 1Q-2DES spectra discussed above, the qualitative characteristics of the 2Q response are highly sensitive to subtle variations in the material structure and composition, reflecting the intricate nature of multi-excitonic correlations. In the lead-based Ruddlesden–Popper perovskites, prior studies combining linear spectroscopy and crystallography have shown that substituting phenylethylammonium (PEA) with fluoro-phenylethylammonium (F-PEA) preserves the excitonic nature while subtly altering exciton–lattice coupling~\cite{koch2025structure}. Nevertheless, Figs.~\ref{fig:2Q_fig}(a) and (b) reveal striking differences in the corresponding 2Q–2DES spectra, suggesting that organic substitution significantly alters multi-excitonic interactions. The 2Q spectrum of \ce{(F-PEA)2PbI4} exhibits four to five well-resolved features, indicative of multiple correlated exciton–exciton pathways, whereas \ce{(PEA)2PbI4} displays a single broad feature near the diagonal. Despite this stark contrast, the extracted biexciton binding energies are comparable, 46\,meV for \ce{(F-PEA)2PbI4} and 50\,meV in \ce{(PEA)2PbI4}, suggesting that the organic substitution influences the coupling landscape and state mixing more strongly than the net exciton–exciton binding strength. 

The 2Q response of \ce{(PEA)2SnI4} is qualitatively similar to \ce{(F-PEA)2PbI4}, albeit with larger noise, and the feature below the diagonal confirms biexciton formation associated with the higher energy excitonic transition ($X_2$). Its biexciton binding energy is comparatively lower than both lead-based structures, with a value of 10\,meV~\cite{rojas2023many}. 
Interestingly, while \ce{(PEA)2SnI4} exhibits a larger exciton–exciton interaction parameter—implying stronger Coulomb coupling—the biexciton binding energy follows the opposite trend relative to \ce{(PEA)2PbI4}. This apparent discrepancy underscores that biexciton binding and exciton–exciton scattering, though both manifestations of many-body Coulomb interactions, arise from distinct microscopic mechanisms: the former depends on attractive or repulsive correlations between charge carriers, while the latter reflects elastic scattering processes governed by dipole strength and the interaction potential~\cite{rojas2023many}. This highlights the deeper motivation for this work, which is understanding the interplay of multi-excitonic interactions and how each one manifests in comparison to the other. Ultimately, 2Q-2DES stands out as the most incisive tool for disentangling these intertwined effects, as it directly maps the coherence pathways linking excitonic and multi-excitonic states, providing a window into the many-body landscape that remains hidden to other spectroscopic approaches.

\section*{Biexcitons in context}

For several decades, excitons in quantum wells (QWs) have served as a model platform for studying many-body effects in low-dimensional semiconductors. In \ce{GaAs}-based QW heterostructures, the valence band splits into heavy-hole and light-hole subbands, resulting in two distinct exciton states: the heavy-hole exciton (HX) and the light-hole exciton (LX), with HX exhibiting a stronger oscillator strength~\cite{bataev2022heavy, deveaud1991enhanced, andreani1991radiative}. The exciton binding energy in GaAs QWs is relatively small, typically only a few meV~\cite{shuvayev2006self, bastard1982exciton}, reflecting the moderate Coulomb interaction in these systems. This rich excitonic landscape adds complexity to biexciton formation, as multiple exciton flavors can interact. 

Early PL data suggest that a biexciton species exists in GaAs QW at low temperature, as a secondary shoulder appears in the emission spectra with increasing pump excitation intensities. The biexciton binding energy extracted from the PL was $\sim$1\,meV, significantly smaller than that of many other 2D materials~\cite{miller1985excitons, miller1982biexcitons}. Further experimental work was done, including polarization-dependent 2D spectroscopy, to determine that there are various types of biexcitons in this class of materials. The optical selection rules for the two excitons (HX and LX) differ with respect to the circular polarization of the light field, and since biexcitons are formed through correlations between opposite-spin excitons, cross- and co-circular polarizations selectively generate distinct biexciton coherences. These findings revealed the existence of both “pure” and “mixed” biexciton species and highlighted how spin and polarization degrees of freedom strongly influence multi-exciton physics in QWs~\cite{stone2009two, turner2010coherent}. As such, GaAs QWs played a foundational role in establishing biexciton spectroscopy as a tool for probing many-body interactions in semiconductors.

Transition metal dichalcogenides (TMDCs), such as \ce{MoS2}, \ce{MoSe2}, \ce{WS2}, and \ce{WSe2}, represent a class of atomically thin semiconductors where excitonic effects are exceptionally strong due to reduced dielectric screening and quantum confinement~\cite{shang2015observation, you2015observation}. Their layered structure, composed of a transition metal plane sandwiched between two chalcogen layers and held together by weak van der Waals forces, supports tightly bound excitons with binding energies of several hundred meV~\cite{chernikov2014exciton, ye2014probing, he2014tightly, ugeda2014giant, you2015observation, yang2022transition}. In contrast to GaAs, where biexciton binding energies are only ~1 meV, TMDCs exhibit values an order of magnitude larger (20–70 meV), allowing biexciton stability even at room temperature~\cite{pandey2019unraveling, lee2016identifying}.

Photoluminescence spectroscopy provided the earliest experimental evidence for biexcitons in TMDCs, with biexciton emission peaks appearing at cryogenic temperatures~\cite{plechinger2015identification}. However, interpretation of the PL spectra is complicated by spectral congestion arising from trions, defect states, and phonon-assisted processes, leading to some debate over biexciton assignments. To resolve this, nonlinear techniques such as pump–probe and two-dimensional coherent spectroscopy have been employed, offering clearer signatures of biexciton formation and enabling precise determination of binding energies~\cite{steinhoff2018biexciton, hao2017neutral}.

An additional layer of complexity in TMDCs arises from the presence of multiple valleys (K and K$^{\prime}$) and strong spin–orbit coupling, which give rise to spin- and valley-polarized excitons. As a result, biexcitons in TMDCs can form in different configurations, such as intra-valley, inter-valley, and spin-forbidden states, each with distinct optical selection rules and coherence properties~\cite{steinhoff2018biexciton, mai2014many, hao2017neutral}. These valley-contrasting biexciton states are not only of fundamental interest but also open avenues for valleytronic and quantum optical applications.

While TMDC monolayers emphasize the impact of reduced dimensionality and valley physics on biexciton formation, another important class of systems where biexcitons have been extensively studied are semiconductor quantum dots (QDs). In these zero-dimensional nanostructures, strong spatial confinement amplifies Coulomb interactions, leading to large and tunable biexciton binding energies. Among them, perovskite QDs have emerged as particularly versatile platforms. 
These nanocrystals, typically based on the \ce{CsPbX3} (X = Cl, Br, I) family, fabricated via colloidal synthesis methods, combine the strong quantum confinement of conventional QDs with the defect tolerance and facile synthesis of halide perovskites~\cite{huang2016colloidal}. They exhibit high PL quantum yields (PLQY), broad bandgap tunability across the visible spectrum (400-700\,nm), and exciton binding energies on the order of hundreds of meV~\cite{zhang2015brightly, ravi2016band, li2018excitons, poonia2021intervalley, barfüsser2024biexcitonic}. Strong Coulomb stabilization leads to biexciton binding energies significantly larger than those found in GaAs QWs and comparable to the values in TMDC monolayers. Understanding biexciton recombination is central to the optoelectronic application of QDS, because exciton dynamics, both single and multi-particle, govern the performance of light-emitting devices.  Biexciton states in perovskite QDs can decay radiatively or through the nonradiative Auger process, and time-resolved PL together with photon-correlation measurements, have been employed to disentangle these pathways~\cite{li2018excitons, makarov2016spectral, eperon2018biexciton}. Multiple studies confirm biexciton–exciton cascades in perovskite QDs, highlighting their utility for deterministic single-photon and entangled-photon generation, and establishing them as promising candidates for quantum information technologies~\cite{lv2019quantum, utzat2019coherent, stevenson2006semiconductor, suzuki2016coherent, huang2020inhomogeneous}.

Unlike GaAs QWs and TMDCs, multidimensional spectroscopy studies of perovskite QDs remain limited. Reported biexciton binding energies from PL and transient absorption (TA) experiments span tens to hundreds of meV~\cite{wang2015all, makarov2016spectral, castañeda2016efficient, ashner2019size, shulenberger2019setting, aneesh2017ultrafast, huang2020inhomogeneous}, reflecting substantial uncertainty. This ambiguity arises largely from dot-to-dot heterogeneity and the relatively broad emission linewidths, which complicate efforts to map the size dependence of biexciton binding energies with precision.

RPMHs occupy a unique position within the landscape of biexciton-hosting materials. They combine features of quantum wells, monolayer semiconductors, and colloidal quantum dots, offering a hybrid platform where strong Coulomb interactions, dielectric confinement, and structural tunability converge. Like TMDC monolayers, 2D perovskites exhibit enhanced excitonic effects due to reduced screening and quantum confinement, often yielding biexciton binding energies significantly larger than those in GaAs QWs. However, unlike atomically thin TMDCs, 2D perovskites consist of inorganic quantum-well layers separated by organic spacers, introducing anisotropic dielectric environments and unique lattice interactions where structural parameters shape their many-body excitonic landscape.

Biexciton binding energies in RPMHs typically range from tens to hundreds of meV, placing them between the low-binding regime of GaAs QWs ($\sim$1 meV) and the tightly bound biexcitons observed in TMDCs and perovskite QDs (20–70 meV and above). This intermediate binding strength supports biexciton stability at elevated temperatures and makes RPMHs promising candidates for room-temperature nonlinear optical applications. Moreover, the presence of multiple excitonic resonances with distinct phonon dressing enables a diverse set of biexciton configurations. While these features may mirror the valley-contrasting biexciton physics in TMDCs, they notably emerge from distinct structural and electronic origins.

\section*{Perspective}
The body of work reviewed here establishes two-dimensional coherent spectroscopy—particularly double-quantum implementations—as the most incisive approach for identifying and quantifying biexciton states in Ruddlesden–Popper metal halides. Across a range of compositions, metal centers, and organic spacers, these techniques consistently reveal biexciton binding energies on the order of tens of millielectronvolts, confirming the robust stabilization of correlated multi-exciton states in these quantum-well derivatives. At the same time, the marked diversity of two-quantum spectral responses observed among nominally similar materials underscores that biexciton physics in RPMHs cannot be captured by a single binding-energy metric alone.

A central conclusion emerging from these studies is that the spectral structure of biexcitons—the number, distribution, and coherence of two-quantum features—is exquisitely sensitive to the underlying material architecture. Subtle changes in organic cation chemistry, metal composition, or lattice softness can dramatically reshape the biexciton landscape without necessarily altering the net exciton–exciton binding strength. This sensitivity reflects the fact that biexcitons form and evolve within a complex, dynamically disordered energy landscape defined by exciton–phonon coupling, dielectric confinement, and local structural heterogeneity.

Looking forward, this dependence positions biexciton spectroscopy as a fundamental probe of exciton quantum dynamics in soft, low-dimensional semiconductors. Rather than serving solely as signatures of many-body Coulomb attraction, biexciton resonances encode how excitons sample, couple to, and remain coherent within fluctuating potential landscapes. In RPMHs—where excitons are increasingly understood as polaronic quasiparticles—the biexciton spectral manifold provides direct access to how lattice dressing, dynamic disorder, and structural tunability shape correlated quantum states.

Future advances will require moving beyond static descriptions of biexciton binding toward a dynamical framework in which biexciton formation, coherence, and decay are explicitly linked to structural fluctuations and phonon-driven disorder. Time-resolved and temperature-dependent multidimensional spectroscopies, combined with theory that treats exciton–phonon and exciton–exciton interactions on equal footing, will be essential for this effort. In this broader context, understanding biexciton spectral structure is not only a question of many-body energetics, but a pathway to uncovering how quantum coherence and correlation persist in complex, dynamically evolving materials—an insight that will be central to the rational design of excitonic and quantum-optical functionalities in halide perovskites and beyond.

\section*{Author contributions}
All co-authors contributed to the redaction of the mini-review manuscript. 

\section*{Conflicts of interest}
There are no conflicts to declare.

\section*{Data availability}
All data presented in this manuscript are reproduced with permission from published sources that are duly cited in the manuscript.
%

\section*{Acknowledgements}
ARSK acknowledges funding from the National Science Foundation CAREER grant (CHE-2338663), start-up funds from Wake Forest University, funding from the Center for Functional Materials at Wake Forest University. CSA acknowledges funding from the Government of Canada (Canada Excellence Research Chair CERC-2022-00055) and support from the Institut Courtois, Facult\'e des arts et des sciences, Universit\'e de Montr\'eal (Chaire de Recherche de l'Institut Courtois). 

\section*{Author Biographies}



\balance

\renewcommand\refname{References}

\bibliography{rsc} 
\bibliographystyle{rsc} 
\end{document}